\documentclass[twocolumn,prb]{revtex4}
\usepackage{amsmath,graphicx,amsfonts,xcolor,float}
\usepackage[normalem]{ulem}
\begin{document}
\author{Trithep Devakul and David A. Huse}
\affiliation{Department of Physics, Princeton University, NJ 08544, USA}
\title{Anderson localization transitions with and without random potentials}
\begin{abstract} 
    We explore single-particle Anderson localization due to nonrandom quasiperiodic potentials in two and three dimensions.  We introduce a class of self-dual models that generalize the one-dimensional Aubry-Andr\'e model to higher dimensions.  In three dimensions (3D) we find that the Anderson localization transitions appear to be in the same universality class as for random potentials.  In scaling or renormalization group terms, this means that randomness of the potential is irrelevant at the Anderson localization transitions in 3D.  
In two dimensions (2D) we also explore the Ando model, which is in the symplectic symmetry class and shows an Anderson localization transition for random potentials.  Here, unlike in 3D, we find that the universality class changes when we instead use a quasiperiodic potential.  
\end{abstract}
\maketitle


\section{Introduction}

Anderson localization of quantum particles is usually discussed in the context of random potentials, as in the original paper of Anderson.\cite{anderson1958}  But the essential physics of Anderson localization does not require randomness:  In a tight-binding lattice model what is needed is that the differences in potential between nearby sites are large enough compared to the hopping matrix elements to ``detune'' most inter-site resonances and thus suppress hopping sufficiently.  These differences in potential (``detunings'') can be due to randomness, but they can alternatively be produced deterministically without any randomness, with quasiperiodic potentials being an example of such a deterministic localizing potential.  The physics of single-particle Anderson localization in such nonrandom quasiperiodic potentials has been much studied in one dimension, particularly for the self-dual Aubry-Andr\'e (AA) chain.\cite{1daa}  Quasiperiodic localization also survives in higher dimensions, as discussed, e.g., in Refs. [\onlinecite{sokoloff,jb}], and in one dimension in the presence of interactions in the form of many-body localization~\cite{qpmbl,ksh,iyer}.
Quasiperiodic potentials are also relatively straightforward to realize experimentally in ultracold atomic experiments.\cite{coldatom1,coldatom2,coldatom3,coldatom4}

One possible perspective on Anderson localization in quasiperiodic potentials is that Anderson localization requires a random potential and quasiperiodic potentials are ``random enough'' to produce localization.  This perspective has lead to labelling the quasiperiodic potential as ``quasi-random'' or ``quasi-disordered''.  Here we instead take a different perspective, namely that quasiperiodic potentials and random potentials are quite different, and one should explore and understand the ways and situations where they behave the same or behave differently.  For us, this perspective arises due to recent work on many-body localization in one dimension, where it is found that added randomness is relevant at the many-body localization phase transition in nonrandom quasiperiodic systems, causing an apparent crossover to a different universality class for random systems.\cite{ksh,qpmbl}


Let's first summarize the situation for the one-dimensional ($d=1$) Aubry-Andr\'e (AA) chain\cite{1daa}, which is a single-particle model described by the Hamiltonian
\begin{equation}
    \mathcal{H}_\text{AA} = \sum_{x} \left(c^\dagger_{x+1} c_x + \text{h.c.}\right) + 2v\sum_x \cos\left(2\pi b x\right) c^\dagger_x c_x~,
    \label{eq:1DAA}
\end{equation}
with $b$ an irrational number (often chosen to be the golden mean).
Upon Fourier transformation, which exchanges real and momentum space, this model can be shown to be dual to itself, with $v\rightarrow v^{-1}$.
Thus, since the eigenstates of this model are localized in real space for large $v$, they are localized in \emph{momentum} space for small $v$ and the particles move ballistically in this small-$v$ phase.
Furthermore, if there is a single localization transition between these two phases it must be at exactly $v_c=1$, where the model is invariant under the duality.
This is indeed the case for the AA model (although there are variations that experience a richer transition~\cite{sarma1,sarma2,sarma3,sarma4,sarma5,selfdual0,selfdual1,selfdual2,selfdual3,selfdual4,sarang}, so self-duality alone does not imply a single transition that occurs simultaneously at all energies).  Now let's consider adding a weak site-random potential to this model.  Such an uncorrelated random potential produces ``hopping'' matrix elements between any two momenta and thus destroys any localization in momentum space, a result that is valid in any dimension.  In one dimension, once we have a random potential all eigenstates are localized, so the distinction between the two phases is removed by the randomness.  At the localization transition we can ask if the randomness is relevant according to the Harris criterion.\cite{harris,ccfs}  It indeed is, since the correlation length exponent $\nu=1<2/d=2$, and thus for this one-dimensional model the randomness is relevant both in the ballistic phase and at the phase transition, and as a result both are destabilized, leaving only the localized phase at nonzero random potential.



In this paper, we generalize the quasiperiodic AA model to more than one dimension, preserving its self-duality.  In our generalization, the self-duality becomes a combination of the Fourier transform, time-reversal, and a relabeling, as we explain below.  Two dimensions appears to be, as is common, a marginal dimension so the behavior is less clear; we defer the exploration of this generalized AA model in two dimensions to future work.\cite{unpublished}  The existence of a ballistic phase in certain 2D quasiperiodic potentials has been recently proven.\cite{cmp,yulia}
We show below that our three-dimensional AA model has an intermediate self-dual diffusive phase between the localized and ballistic phases, with mobility edges.  Considering the stability to added randomness, the ballistic phase and thus the ballistic-to-diffusive phase transition are unstable to randomness.  
But the 3D orthogonal Anderson localization transition has correlation length exponent $\nu_{3D}\cong 1.6>2/d=2/3$, so the Harris criterion says randomness is irrelevant at the diffusive-to-localized transition. 
The quantum kicked rotor with three incommensurate frequencies has also already been observed\cite{kickedrotor} to exhibit the same critical phenomena as the 3D orthogonal Anderson localization transition.
Thus randomness is irrelevant, in the renormalization group meaning of this word, for Anderson localization in $d=3$.  We find that this is also true for our self-dual 3D nonrandom quasiperiodic model, which has a localization transition that appears to be in the same universality class as that with randomness.

This generalized AA model has a time-reversal invariant Hamiltonian and spinless particles, so is in the orthogonal symmetry class, which does not exhibit an Anderson localization phase transition with randomness in two dimensions.\cite{andersonreview}  To explore these issues a bit more, we also examine a non-self-dual localization model  in the symplectic universality class (the Ando model) with tunable randomness, which does show an Anderson localization phase transition in two dimensions with a random potential.  We find that in three dimensions, for this symplectic model the transition has the same apparent exponents for nonrandom quasiperiodic potentials as it has for random potentials, as in the orthogonal class 3D models.  
In the two-dimensional symplectic Ando model, however, the localization transition for the nonrandom quasiperiodic model appears to be in a quite different universality class from the same model with random potentials.

\section{Generalized Aubry-Andr\'e model}
Our generalization of the AA model to higher dimension $d$ is described by the Hamiltonian
\begin{equation}
    \mathcal{H} = \sum_{\vec{r}}\sum_{i=1}^d \left( c_{\vec{r}+\hat{u}_i}^\dagger c_{\vec{r}}+\text{h.c.} \right) + \sum_{\vec{r}} V(\vec{r}) c_{\vec{r}}^\dagger c_r
    \label{eq:Ham1}
\end{equation}
where the sum is over all sites $\vec{r}$ on a hypercubic lattice (with lattice constant unity),  $\hat{u}_i$ are the lattice basis vectors, and $c_{\vec{r}}^\dagger, c_{\vec{r}}$ are the usual particle creation and annihilation operators.  In this paper we will focus on three dimensions, but for many aspects the generalization to all $d$ is straightforward, so we will present general $d$ before restricting to $d=3$.
$V(\vec{r})$ is the onsite potential term, which is given by
\begin{equation}
    V(\vec{r}) = 2v\sum_{i=1}^{d} \cos(2\pi\vec{b}_{i}\cdot \vec{r} + \phi_i)
    \label{eq:V}
\end{equation}
where $v$ is a constant used to control the strength of the potential, and $\left\{\vec{b}_i\right\}$ are $d$ linearly independent vectors which together determine vital characteristics of the resulting model.
The $\phi_i$ are arbitrary phases in the cosines, which will simply amount to an overall translation in the thermodynamic limit.
It is convenient to refer to the matrix $\mathbf{B}$ whose elements $B_{i j}$ are the $j$th component of the $\vec{b}_i$.
First, notice from eq.~(\ref{eq:V}) that since $\vec{r}$ has integer components, each element $B_{i j}$ only has meaning modulo 1.
We also want the potential to be quasiperiodic, and not periodic in any sense: one way to impose this condition is to require that $(\vec{b}_{i}\cdot \vec{r})$ is nonzero and irrational for all $i$ and all lattice vectors $\vec{r}$.  This requirement also prevents the model from being separable in to two or more lower-dimensional models.
We will choose to focus on a particular subset of possible $B_{i j}$, as will be discussed in the next section.  

This model, similar to its lower-dimensional counterparts, can be thought of as a Fourier slice of a quantum Hall state in a system of twice the spatial dimension~\cite{2dqh,4dqh}.
In this case, there is a corresponding ($2d$)-dimensional model of a particle hopping in a particular magnetic field configuration that, upon picking a single Fourier component of each of $d$ dimensions, reduces to our model.  
In one and two dimensions the spectrum of this model contains energy gaps of varying sizes, and due to the topological nature of the quantum Hall state, opening the boundary creates edge-localized states that traverse the gaps as a function of $\phi_i$.\cite{2dqh,unpublished}

\subsection{Self-duality}
We will now construct finite-system-size approximations of the Hamiltonian~(\ref{eq:Ham1}) with periodic boundary conditions, and determine the necessary conditions for an exact AA-type self-duality.  This then naturally produces a self-dual quasiperiodic system in the limit of an infinite system.

Consider the model defined on a finite $L\times L \times ... \times L$ hypercubic lattice with periodic boundary conditions.
The periodicity condition $V(\vec{r}) = V(\vec{r}+L\hat{u}_i)$ requires that $L B_{i j} \in \mathbb{Z}$.
Furthermore, due to the finite size of the system, adding an arbitrary phase to the cosine terms in $V(\vec{r})$ no longer results in an overall translation.  
One can now also introduce magnetic fluxes threading each cycle of this hypertorus, resulting in an overall phase factor gained by a particle hopping around the cycle in each of the directions.

We first present the exactly self-dual Hamiltonian, and present justification for each of its features after.  
The self-dual version of the Hamiltonian is given by
\begin{equation}
    \mathcal{H}_\text{SD} = \sum_{\vec{r}}\sum_{i=1}^d \left(e^{i\phi_i} c_{\vec{r}+\hat{u}_i}^\dagger c_{\vec{r}} + \text{h.c.} \right) +\
    \sum_{\vec{r}} V_\text{SD}(\vec{r}) c_{\vec{r}}^\dagger c_{\vec{r}}
    \label{eq:HamSD}
\end{equation}
with 
\begin{equation}
    V_\text{SD}(\vec{r}) = 2v\sum_{i=1}^{d} \cos\left(2 \pi \sum_{j=1}^{d} B^{\text{SD}}_{i j}r_j + \phi_i\right)
    \label{eq:VSD}
\end{equation}
where $\mathbf{B}^\text{SD}$ is a matrix satisfying
\begin{eqnarray}
    L B^\text{SD}_{i j} = L B^\text{SD}_{j i} \in \mathbb{Z}_L\label{eq:sdcond1} \\ 
    \gcd(\det\left( L \mathbf{B}^\text{SD}\right), L) = 1 \label{eq:sdcond2}
\end{eqnarray}

The self-duality transformation consists of three steps:
\begin{enumerate}
    \item Time reversal. The signs of all the magnetic fluxes are reversed.
    \item Fourier transformation.  We replace real space operators by momentum space operators in the standard way, $c_{\vec{k}}^{\dagger} = {L^{-d/2}}\sum_{\vec{r}}e^{i \vec{k}\cdot\vec{r}} c_{\vec{r}}^\dagger$.  This switches the role of the potential and kinetic terms.
    \item Relabeling of momenta. To get the resulting Hamiltonian back to a recognizeable form, we relabel the momenta as $\vec{k}\rightarrow L\mathbf{B}^\text{SD} \vec{k}$.  
\end{enumerate}
The result of these transformations is an identical Hamiltonian except in terms of momentum space operators, and the coefficient $v$ has been moved to the hopping term.
Representing the Hamiltonian matrix in the single-particle subspace, the first step is complex conjugation and the second and third step can be encoded as a unitary transformation.

The Fourier transformation switches the role of the kinetic and potential term, and so the fluxes and phase shifts also switch roles.  
However, due to the asymmetry of the Fourier transform, the phase shifts after the transformation have all picked up a minus sign.
Thus, this sign flip is cancelled out by first performing a time reversal.

We now comment on the restrictions to $\mathbf{B}^\text{SD}$.  
The elements $L B_{i j}^\text{SD}$ are integers mod $L$ due to the periodicity requirement.
The symmetry restriction is due to the fact that after the self-duality transformation, the resulting momentum space Hamiltonian contains the transposed matrix.
Finally, the need for last restriction (Eq~\ref{eq:sdcond2}), the coprimality of $\det(L \mathbf{B}^\text{SD})$ with $L$, arises in the relabeling step.
In order for the relabeling to be an isomorphism, the transformation must be invertible.
This implies that $\det (L\mathbf{B}^\text{SD})$ must be invertible mod $L$, and therefore coprime to $L$.
It is interesting to note that since the flux is only well defined modulo $2\pi/L$, so must the phase shifts, and therefore shifting $\phi_i \rightarrow \phi_i + 2\pi/L$ must result in an overall translation of the origin to some other point.  
This is non-obvious, but follows from the the coprimality condition.

If we consider an infinite model with some symmetric quasiperiodic $\mathbf{B}$, then one can make a series of periodic self-dual approximants to this model for each $L$ by choosing the rational 
$\mathbf{B}^\text{SD}$ that satisfies the above conditions and is closest to $\mathbf{B}$.  Any such sequence defines a self-dual model in the thermodynamic limit $L\rightarrow\infty$.

\subsection{Specific model details}\label{ssec:spec}
We now focus on a particular form for $\mathbf{B}$ defined for the infinite three-dimensional system. 
We choose an explicitly symmetric orthonormal matrix $\mathbf{R}(\theta)$ whose action can be thought of as a reflection followed by three Euler rotations by an angle $\theta$.  
Letting $c=\cos\theta$ and $s=\sin\theta$, this matrix takes the form
\begin{equation}
    \mathbf{R}(\theta) = 
    \begin{bmatrix}
        c^2+s^3 & c s & c s^2 - c s \\
        c s & -s & c^2 \\
        c s^2 - c s & c^2 & c^2 s + s^2
    \end{bmatrix}
    \label{eq:Rmat}
\end{equation}
We then take $\mathbf{B} = b \mathbf{R}(\theta)$, where $b$ is a constant.
As long as one avoids $\theta = 0$, $\theta=\pi/2$ and multiples thereof, this will result in a non-separable model.
It is also aesthetically pleasing as all three vectors $\vec{b}_i$ are now orthogonal with equal magnitude.

One can construct exactly self-dual approximations of this model on a finite system of size $L^3$ by picking the closest integer approximation $L \mathbf{B}^\text{SD} \approx L\mathbf{B}$.
However, one may occasionally run into the issue where $\gcd(\det(L\mathbf{B}),L)\neq 1$.  
This is typically easily remedied by picking a slightly different integer approximation (instead of rounding a value down, round it up instead), or if that fails, picking a slightly different $L$.
For prime $L$, one can almost certainly find such an approximation that works.
In this way, it is possible to take a series of self-dual models to the thermodynamic limit $L\rightarrow\infty$, which is therefore also self-dual.

The choice of $b$ and $\theta$ turn out to be important when exploring finite systems numerically.  
For instance, consider the seemingly innocuous choice of $b=(\sqrt{5}-1)/2\approx0.618$ (the inverse golden ratio) and $\theta=\pi/8$.
The resulting matrix actually has the property that for the lattice vector $\vec{w}=(2,0,-3)$, $\mathbf{B}\vec{w} \approx (0.0029, 0.0009, -0.0027) \mod 1$.  
Thus, the potential $V(\vec{r})\approx V(\vec{r}+\vec{w})$ turns out to be almost periodic!  
This introduces a large additional lengthscale into the problem, as the localization length along the $\vec{w}$ direction will be large, even well in to the localized phase.
Thankfully, such an occurence turns out to be rare, and for most choices of $b$ and $\theta$ such coincidences do not occur with such a small $\vec{w}$. 
For the remainder of the paper, we will focus on the particular choice $b=(\sqrt{5}-1)/2$ and $\theta=\pi/7$ for this 3D model, which is far more typical and does not exhibit such a coincidence.  
We have also looked at other typical choices of $b$ and $\theta$, which yield similar results.

We will first present results for such self-dual approximations, demonstrating the existence of an intermediate diffusive phase between the real-space localized and momentum-space localized phases, in which eigenstates are delocalized in both real and momentum space.
These phases can be easily distinguished by their dynamics, which go from ballistic at low $v$, to diffusive in the intermediate phase, and finally localized at high $v$.
Then, we will open the boundary conditions and study the finite-size scaling behavior of the localization phase transition using the tools of multifractal analysis, where we find the critical scaling exponent of the correlation length $\nu=1.6\pm0.1$, which is in agreement with the accepted value of this critical exponent for the (3D GOE) Anderson localization transition with random potentials, $\nu\approx1.57\pm0.01$.~\cite{slevin1}
This indicates that this transition exhibits the same universality class for nonrandom quasiperiodic potentials as it does with random potentials.

\section{3D Results}
\subsection{Intermediate phase and mobility edges}

\begin{figure}[t]
    \centering
    \includegraphics{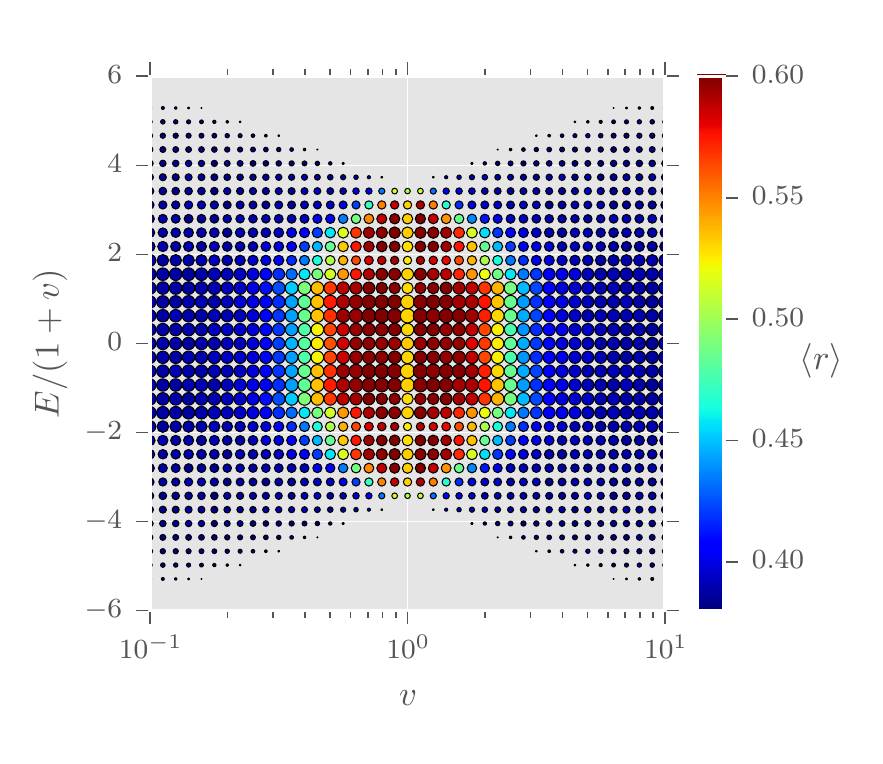}    \caption{The value of the mean level spacing ratio $\langle r \rangle$ (defined in the text) as a function of quasiperiodic potential strength $v$ and energy $E$ for the exactly self-dual 3D model for $L=20$ with periodic boundary conditions.
        The size of the points is proportional to the density of states.
        For high or low $v$ the model is localized in real or momentum space, respectively, and so the level spacing ratio shows Poisson statistics (${\langle r \rangle }_\text{Pois} \approx0.3863$).  
        An intermediate phase exists between these two limits, where the level statistics show GUE behavior (${\langle r \rangle }_\text{GUE}\approx 0.5996$) for $v\neq 1$, and GOE behavior (${\langle r \rangle}_\text{GOE}  \approx0.5307$) at exactly $v=1$ when the system is time-reversal symmetric.  
        This indicates that the eigenstates are delocalized in both real and momentum space within this intermediate phase.
        Also notice that the spectrum is symmetric about $E=0$: this is because a simultaneous phase shift $\phi_i \rightarrow \phi_i+\pi$ of all three phases flips the sign of the entire Hamiltonian.
    }\label{fig:rcolor}
\end{figure}

\begin{figure}[t]
    \centering
    \includegraphics[width=0.5\textwidth]{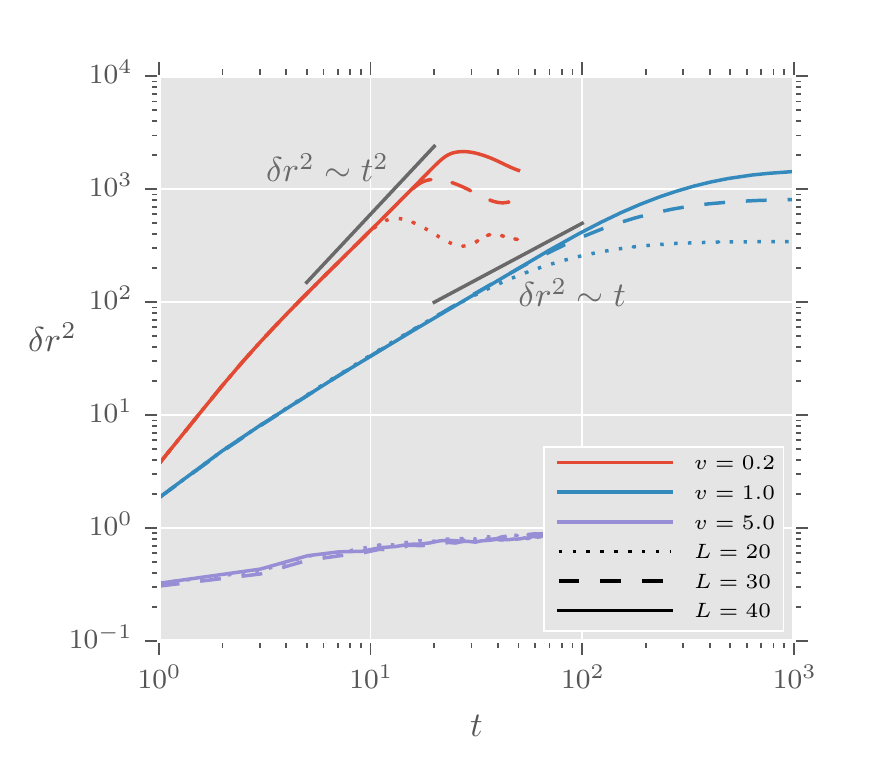}    
    \caption{\
        The growth with time of the mean-square displacement of a particle initially placed at the corner of an open 3D system, for $v=0.2,1,5$, which are well within each of the three phases, which show ballistic, diffusive, and localized behavior respectively.  Note that the former two also show finite-size saturation at long times when they spread over the full sample.
    }\label{fig:r2dynamics}
\end{figure}

We first perform exact diagonalization on the exactly self-dual approximation of the previously described model with $b=(\sqrt{5}-1)/2$ and $\theta=\pi/7$.
To identify the phases of the model, we look at the level spacing ratios, defined for the $n$th eigenstate as $r_n = \min(\delta_n,\delta_{n+1})/\max(\delta_n,\delta_{n+1})$, where $\delta_n=E_{n}-E_{n-1}$ and $E_n$ are the sorted energy eigenvalues in increasing order.
In well-localized phases (in either real or momentum space), neighboring eigenvalues do not repel, and thus the level spacings are distributed according to a Poisson distribution, with a characteristic mean value of $\langle r \rangle$.  
When the eigenstates are delocalized, neighboring eigenvalues repel and $\langle r \rangle$ takes on the characteristic GUE or GOE value with or without time-reversal symmetry, respectively.

The self-dual Hamiltonian~(\ref{eq:HamSD}) can be shown to be time-reversal symmetric at the self-dual point $v=1$ for any choice of phases $\phi_i$, despite being complex. 
This is because the self-duality transformation consists of a complex conjugation step followed by a unitary transformation.
Thus, at $v=1$, the complex conjugate of the Hamiltonian is still unitarily related to itself, $\mathcal{H} = U \mathcal{H}^{*} U^\dagger$.
This shows that there is a basis in which the Hamiltonian is real, so it should obey GOE level statistics when delocalized.  Away from $v=1$ the model does not
have time-reversal symmetry, except for the special case where all three phases $\phi_i$ are exactly either zero or $\pi/L$.

Figure~\ref{fig:rcolor} shows the mean value of $\langle r \rangle$ sampled over eigenstates within energy intervals and over choices of $\phi_i\in\left[0,2\pi/L\right]$.
Notice that there are distinctly three phases. 
For low $v$, the eigenstates are localized in momentum space and show Poisson statistics.
As $v$ is increased, starting near $E=0$ the level spacings begin to show GUE statistics.  
This indicates that the eigenstates are delocalized in both real and momentum space.
At exactly the self-dual point $v=1$, the level spacings show GOE statistics due to time-reversal symmetry, but no sign of a localization phase transition.
For high $v$, the eigenstates become localized in real space and the level spacings once again show Poisson statistics.  The phase boundaries $v_c(E)$ do depend on energy, so this model does exhibit mobility edges where one can cross the transition at fixed $v$ by varying the energy.

These three phases can be readily characterized by their dynamics.
To do this, we consider the system with open boundaries (no hopping across the boundaries) and the particle initially on the corner site $\vec{r}_0=(1,1,1)$.  
Then, one can ask how far the particle has spread after a time $t$, measured by $\delta r^2(t) = \langle \left|\vec{r}(t) -\vec{r}_0\right|^2\rangle$.
This is shown in Fig.~\ref{fig:r2dynamics}, where time evolution of the initial state has been performed using sparse matrices.
At low $v$, eigenstates are localized in momentum space and so the particle spreads ballistically $\delta r^2(t) \sim t^2$.
In the intermediate phase, we find that $\delta r^2 \sim t$, which corresponds to diffusive spreading.
Meanwhile, at high $v$, the particle is localized and so $\delta r^2(t) \rightarrow \text{const}$ at long times.
Note that there appears to be a weak correction to the diffusive spreading at $v=1$ (this weak correction is {\it not} clearly visible in Fig.~\ref{fig:r2dynamics}), which arises from states near the edges of the band.  If we first project onto states near the center of the band before performing the dynamics, this correction is greatly reduced.

\subsection{Critical scaling}

In order to study the scaling at the phase transition, we open the boundary conditions and study finite-sized $L\times L\times L$ systems of the Hamiltonian~(\ref{eq:Ham1}) with the exact quasiperiodic $\mathbf{B}$ matrix described in Sec.~\ref{ssec:spec}.
The reason for this is that the self-dual approximations are slightly different for every choice of $L$, which makes it difficult to analyze finite-size scaling behavior with $L$, especially since small changes in $\mathbf{B}$ can have a significant effect (c.f.\ the model with $\theta=\pi/8$).
Opening boundary conditions also has the advantage that fluxes can be gauged out, and so the Hamiltonian is completely real.
Because the critical point $v_c(E)$ is dependent on energy, we also focus our studies on states at a particular energy.  Note that this localization transition is in the orthogonal symmetry class; the GUE level statistics seen in the finite systems with periodic boundary conditions are only a finite-size effect due to the ``twisted'' boundary conditions.  In the ``bulk'' this model is time-reversal invariant.

While studying the finite-size scaling of the level spacing ratio $\langle r \rangle$ is possible, it is not optimal because it relies on obtaining many consecutive eigenvalues of the $L^3\times L^3$ Hamiltonian matrix, and it does not exploit all the information present in the exact eigenfunctions.
Another method that has proven successful in characterizing the critical properties of the Anderson transition are transfer matrix methods~\cite{slevin1}, in which one observes the  quasi-1D localization length in a $L_x\times L_y \times L$ system as $L\rightarrow\infty$.  
However, the extent to which this method can be applied to the current model is unclear, as the model is not isotropic.
That is, particles may delocalize along a certain (non-lattice) direction more quickly than along other directions, whereas the transfer matrix method only sees one direction at a time.

We therefore examine the transition using a ``multifractal'' finite-size scaling scaling analysis~\cite{multifrac1,multifrac2,multifrac3}.
Note this analysis can still be applied to any localization transition, even if we do not know a priori that critical wavefunctions in this model exhibit multifractal behavior.
Using shift-invert Lanczos, we obtain the eigenstate $\left|\psi\right\rangle$ 
closest to a particular energy.
We then divide the lattice up into boxes of size $\ell^3$ sites, of which there are ${(L/\ell)}^{3} \equiv \lambda^{-3}$ such boxes.
The coarse-grained weight of the wavefunction in the box $k$ is then given by
    $\mu_{k} = \sum_{i \in k} {\left|\left\langle i | \psi\right\rangle\right|}^2$,
where the sum is over all sites $i$ in the box $k$.
The probability distribution of $\mu_k$ can tell us a lot about the properties of $\left|\psi\right\rangle$ at various lengthscales.  
For instance, the inverse participation ratio (IPR) is related to the second moment of this distribution for $\lambda=1/L$ (no coarse-graining).
The reason for coarse graining is that one can now fix $\lambda$ and reasonably compare systems of different size.
In particular, the probability distribution of $\tilde{\alpha}\equiv \log\mu_k / \log \lambda$ at fixed $\lambda$ becomes independent of system size at the critical point.  
Away from the critical point, this distribution drifts in different directions with $L$ in the localized versus the delocalized phase.
Thus the mean value of this distribution, denoted $\tilde{\alpha}_0 = \langle \log \mu \rangle / \log \lambda$, is suitable to be used as a dimensionless finite-size scaling observable.
Although there are many interesting avenues one could take in characterizing the multifractality of the critical states, we stop here and simply take advantage of this finite-size scaling.

\begin{figure}[h]
    \centering
    \includegraphics[width=0.5\textwidth]{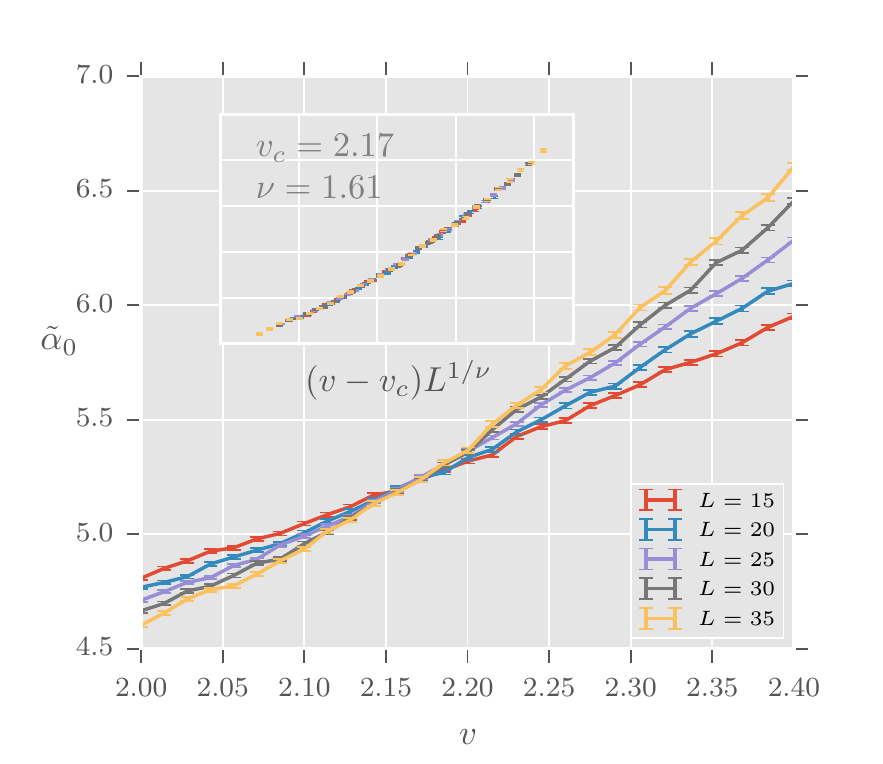}    
    \caption{\
        The quantity $\tilde{\alpha}_0$ (described in the text) is shown for various 3D system sizes near the critical point at $E=1$.  
        The transition is becoming sharper with increasing system size, as expected at a phase transition.  
        A standard finite-size scaling estimate for the critical point and exponent gives a scaling exponent estimate $\nu = 1.6 \pm 0.1$.  
        The scaling collapse is shown in the inset.
    }\label{fig:alphascaling}
\end{figure}

We focus on the scaling of $\tilde{\alpha}_0$ as a function of $L$ and $v$, with $\lambda=0.2$, allowing us to increment $L$ in steps of 5.
We sample over choices of $\phi_i\in\left[0,2\pi\right]$, which amounts to sampling over different positions in the infinite system.  
From each sample the one eigenstate closest in energy to $E=1$ is obtained.
Figure~\ref{fig:alphascaling} shows $\tilde{\alpha}_0(v)$ near the transition for various system sizes $L$, which shows the expected crossings as the functions $\tilde{\alpha}_0(v)$ get ``steeper" as $L$ is increased.
Assuming the standard simple scaling form $\tilde{\alpha}_0(L,v) = f((v-v_c)L^{1/\nu})$ for some universal function $f$ already allows for an excellent collapse of all the data, as shown in the inset of Fig.~\ref{fig:alphascaling}.
This leads to an estimate for the critical scaling exponent of $\nu = 1.6 \pm 0.1$.
Note that by self-duality of our model, the ballistic-to-diffusive transition therefore \emph{also} exhibits the same critical behavior.
This exponent estimate is consistent with the 3D orthogonal class critical exponent $\nu=1.57\pm 0.01$ for the Anderson localization transition with randomness\cite{slevin1}.  
This shows our model behaves as the 3D kicked rotor model\cite{kickedrotor}, indicating that the universality class of the 3D Anderson localization transition is the same for quasiperiodic and for random potentials.

This being the same universality class as the disorder-driven Anderson localization transition therefore implies that this transition is stable to disorder.
One can imagine adding a small amount of disorder to this model by allowing a site-dependent phase shift in the quasiperiodic potentials picked uniformly within a range $\left[-w,w\right]$.
Any $w$ will produce hopping between any two momenta, and therefore completely destroy the ballistic phase.
When $w=\pi$ this becomes the totally random Anderson model with the same potential distribution.
As we know from Anderson localization, in one and two dimensions, any small disorder is relevant and a small $w$ will localize the system in the thermodynamic limit and remove the localization transition that is present in the AA model.
Our result shows that this ceases to be the case in three dimensions, and disorder is in fact irrelevant at this phase transition.

\section{Symplectic symmetry class}

The models studied in the previous two sections of this paper are in the orthogonal symmetry class, where an Anderson localization transition does not exist for two-dimensional (2D) systems with random potentials.  
However, 
there are other symmetry classes for which an Anderson localization transition does exist in 2D random potentials.\cite{andersonreview}
We will now focus on the symplectic symmetry class, in particular, since it is known to have an Anderson localization transition in 2D in the presence of randomness.

We examine the Ando model~\cite{ando}, defined by
\begin{equation}
    \mathcal{H} = \sum_{\vec{r},\sigma,\sigma\prime} \sum_{i=1}^{d} \left( t^{i}_{\sigma \sigma\prime} c^\dagger_{\vec{r}+\hat{u}_i, \sigma} c_{\vec{r},\sigma\prime} + \text{h.c.} \right) + \sum_{\vec{r},\sigma} V_{w}(\vec{r}) c^\dagger_{\vec{r},\sigma} c_{\vec{r},\sigma}
    \label{eq:sympham}
\end{equation}
where $\sigma,\sigma\prime\in\left\{\uparrow,\downarrow\right\}$ labels the $z$-component of the spin-$1/2$ fermion and $d=2$.
The main difference from our previously studied model is that there is now a spin- and direction-dependent hopping.
We have chosen the hopping matrix to be $t^{i}_{\sigma\sigma\prime} = \left(e^{i \frac{\pi}{6} \sigma_i}\right)_{\sigma\sigma\prime}$
where $\sigma_i$ are Pauli matrices $\sigma_x$,$\sigma_y$.
The parameter $w$ can be used to tune randomness in the potential $V_{w}(\vec{r})$, 
\begin{eqnarray}
    V_{w}(\vec{r}) &=&  2v\sum_{i=1}^{d} \cos(2\pi\vec{b}_{i}\cdot \vec{r}+\phi_i + \delta\phi_{i,\vec{r}}) 
    \label{eq:Vw}
\end{eqnarray}
where $\delta\phi_{i,\vec{r}} \in \left[-w,w\right]$ are independent random variables chosen from a uniform distribution, and $v$ controls the overall strength of the potential as before.
The $\left\{\vec{b}_i\right\}$ are chosen to be two orthogonal vectors of magnitude $\left|\vec{b}_i\right|=(\sqrt{5}-1)/2$ rotated $\pi/8$ from the lattice direction.
At $w=0$ this is the quasiperiodic potential in Eq.~(\ref{eq:V}), and at $w=\pi$ this corresponds to a fully random potential, with $w$ in between being ``semi-random'' correlated disorder.

Note that this model no longer has the AA self-duality.  
In Appendix~\ref{app:sympapp}, we have verified that the $w=0$ model in 3D, like with the orthogonal case, 
has a localization transition with a scaling exponent in good agreement with the random Anderson symplectic 3D universality class~\cite{symp3d1,symp3d2}.
In both 2D and 3D, the phase boundaries are no longer symmetric as in Fig. 1, due to this symplectic model's lack of self-duality.

\begin{figure}[h]
    \centering
    \includegraphics[width=0.5\textwidth]{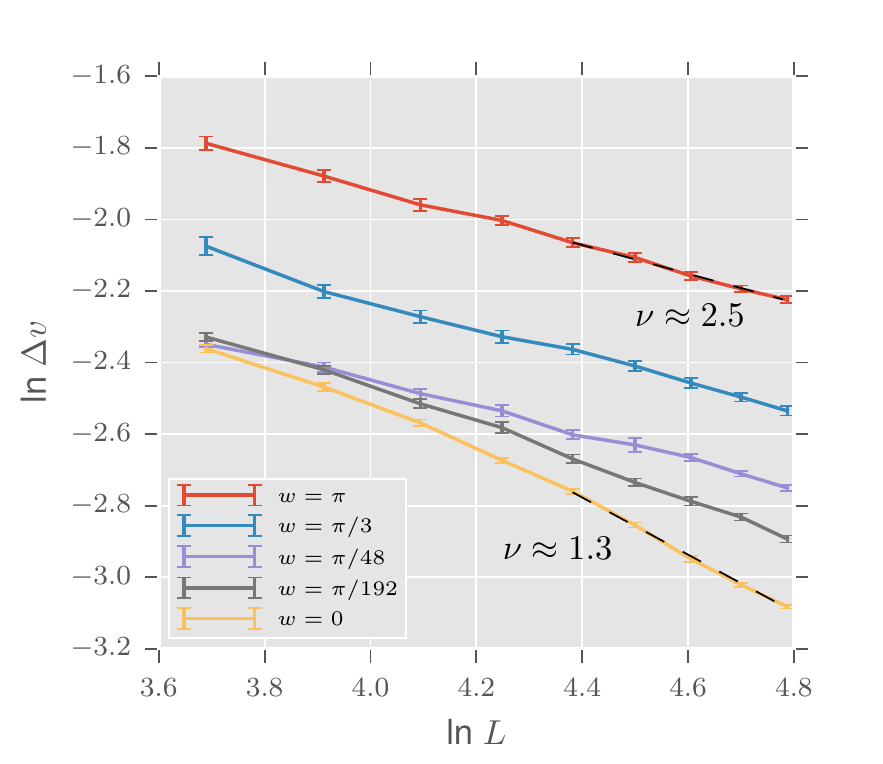}    
    \caption{\
        The transition width $\Delta v$ as a function of $L$ for the 2D symplectic model at energy $E=0$, at varying levels of randomness from quasiperiodic ($w=0$) to fully random ($w=\pi$).  
        The width is defined as the difference in $v$ between when the scaling quantity $\tilde{\alpha}_0$ crosses $2.2$ and $2.4$, which is obtained from a polynomial fit to the data near the transition.
        The width should scale as $\ln \Delta v \sim -(1/\nu) \ln L$ for large $L$.
    }\label{fig:sympwidth}
\end{figure}

We now focus on the critical scaling of the quantity $\tilde{\alpha}_0$ in this 2D Ando model.
Here, we have divided each system into 100 boxes each of size $(L/10)\times (L/10)$ sites (corresponding to $\lambda=0.1$), and computed $\tilde{\alpha}_0$ using only boxes not touching the edge of the system, to avoid any possible effects from edge-localized states.~\footnote{In 3D we kept the edge boxes, since not doing so would have meant keeping only 27 of the 125 boxes there, so would substantially increase the statistical uncertainties.}
We find it instructive to examine the transition width $\Delta v$, which we define as the difference in $v$ between when the curve crosses $\tilde{\alpha}_0 = 2.2$ and $\tilde{\alpha}_0=2.4$.  
The scaling data for the random and quasiperiodic case are shown in Appendix~\ref{app:sympapp}.
This width should scale as $\Delta v \sim L^{-1/\nu}$ for $L\rightarrow\infty$, and is shown in Figure~\ref{fig:sympwidth} for a few choices of $w$.

From a glance, it is immediately obvious that the apparent exponents at large $L$ are different between the quasiperiodic ($w=0$) and random ($w=\pi$) cases.
The random model shows a final slope corresponding to $\nu\cong2.5\pm 0.2$, which is close to results from other methods that have obtained the exponent for the 2D symplectic transition to high precision $\nu\cong2.75\pm0.01$.\cite{symp2d1,symp2d2,symp2d3}
However, the quasiperiodic model may not yet even be in its asymptotic regime, as there is a clear downwards curvature.
The large-$L$ slope corresponds to  $\nu\cong1.3\pm 0.1$, significantly lower than in the random case.
Furthermore, the scaling behavior appears to be very sensitively dependent on $w$.
These finite-size scaling results strongly suggest that the universality class is different between the nonrandom ($w=0$) quasiperiodic and fully random ($w=\pi$) symplectic models in 2D.  Some further data and speculations about this model's behavior appear in the Appendix.

\section{Conclusion}
We have constructed a family of higher-dimensional quasiperiodic localization models as natural generalizations of the 1D AA chain.  
Our generalization can be made to preserve the self-duality from the 1D model, which can be made exact on finite periodic lattices via 
a prescription which we have described in detail. 

We numerically investigated one typical such 3D generalized AA model.
In contrast to the AA chain, there is not a single localization transition at the self dual point $v=1$.
Instead, we clearly demonstrate the existence of an intermediate diffusive phase between the ballistic and localized phases, in which eigenstates are delocalized in both real and momentum space.  
This phase diagram shows mobility edges.
A finite-size scaling analysis of the Anderson localization transition gives a correlation length exponent $\nu=1.6\pm 0.1$, consistent with that of the 3D orthogonal Anderson localization universality class, which therefore does not require randomness.

We then examined the Ando model, which is in the symplectic symmetry class and does possess an Anderson localization transition in 2D.
In 3D this model behaves similarly to the orthogonal class, with randomness in the potential being irrelevant at the nonrandom quasiperiodic localization transition.
But in 2D this Ando model exhibits apparent scaling exponents that are quite different between quasiperiodic  and random potentials.  
This indicates that the localization transition in the 2D symplectic class has at least two different universality classes, one for the nonrandom quasiperiodic model, and a different one for random potentials.

We started this investigation to see if a single-particle Anderson localization model might show behavior analogous to the 1D many-body localization (MBL) transition, where added randomness is strongly relevant at the quasiperiodic transition and produces a crossover to a random universality class.\cite{ksh,qpmbl}  Instead, we have found other scenarios for single-particle localization: Randomness can destroy both the delocalized phase and the transition, as in 1D and for the orthogonal class in 2D.  Randomness can be irrelevant so there is only one universality class, which appears to be the case in 3D and presumably in higher dimensions.  The 2D symplectic case seems to be the closest to the 1D MBL systems in its behavior, with the quasiperiodic transition showing quite different scaling from the random case, but there are also differences from the 1D MBL systems, such as the presence of a ballistic phase in the single-particle quasiperiodic models.



\section{Acknowledgements}   We thank Sarang Gopalakrishnan, Vedika Khemani, Jed Pixley, Ulrich Schneider, Tom Spencer, Attila Szabo and Justin Wilson for helpful discussions.

\newpage
\appendix
\section{Symplectic scaling data}\label{app:sympapp}
Here, we show scaling data for the 3D (quasiperiodic) and 2D (quasiperiodic and random) Ando model.

\begin{figure}[h]
    \centering
    \includegraphics[width=0.5\textwidth]{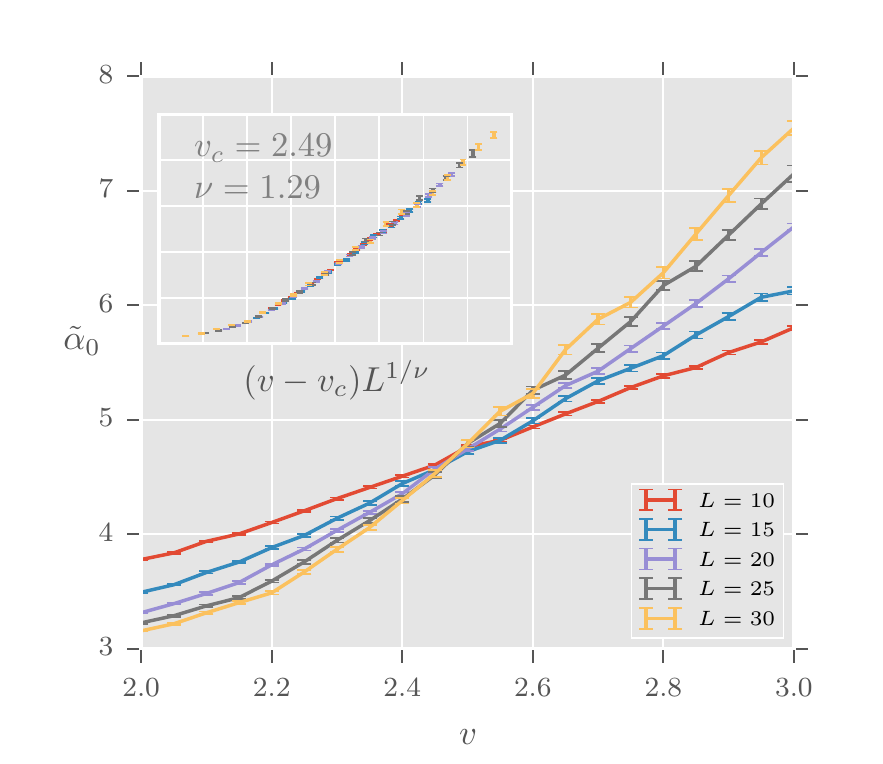}%
    \caption{\
        The scaling of $\tilde{\alpha}_0$ near the transition for the quasiperiodic 3D symplectic model at energy $E=0$.
        Inset shows the scaling collapse.
    }\label{fig:3dsympscaling}
\end{figure}

The 3D Ando model is given by Eq.~\ref{eq:sympham} with $d=3$, with the same hopping matrix $t^{i}_{\sigma\sigma\prime} = \left(e^{i \frac{\pi}{6} \sigma_i}\right)_{\sigma\sigma\prime}$, except that now we use three Pauli matrices $\sigma_{x,y,z}$, with the same potential as the 3D AA model (Eq.~\ref{eq:V} with $\mathbf{B} = b\mathbf{R}(\theta)$ for $b=(\sqrt{5}-1)/2$ and $\theta=\pi/7$).
The scaling of $\tilde{\alpha}_0$ for this model is shown in Fig.~\ref{fig:3dsympscaling}, which shows an exponent $\nu=1.3\pm0.1$, in agreement with the 3D symplectic Anderson value for random potentials of $\nu=1.38\pm0.02$.~\cite{symp3d1}

\begin{figure}[h]
    \centering
    \includegraphics[width=0.5\textwidth]{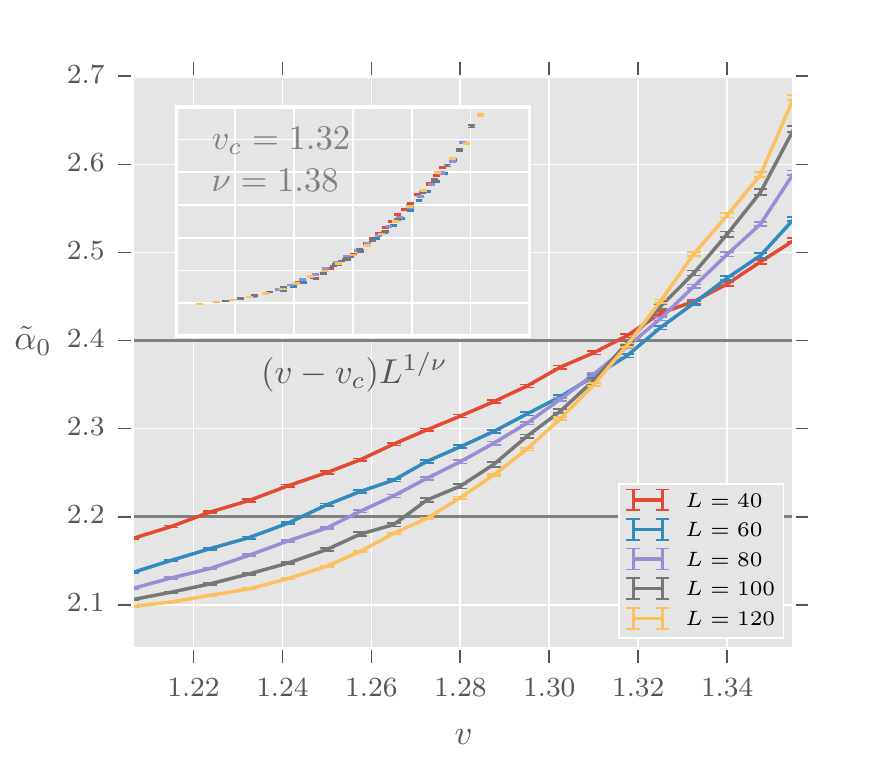}%
    \caption{\
        The scaling of $\tilde{\alpha}_0$ near the transition for the quasiperiodic ($w=0$) 2D symplectic model (Eq.~\ref{eq:sympham}) at energy $E=0$.
        The width of the transition $\Delta v$ is defined by where the curves cross the grey lines.
        Inset shows the scaling collapse.
    }\label{fig:qpsympscaling}
\end{figure}

\begin{figure}[h]
    \centering
    \includegraphics[width=0.5\textwidth]{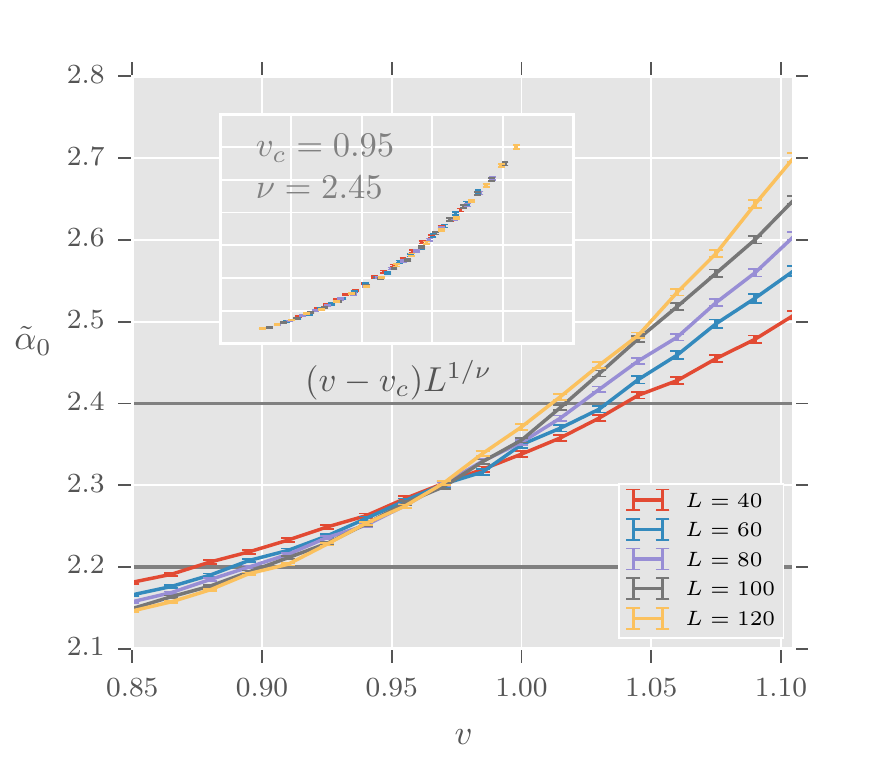}%
    \caption{\
        The scaling of $\tilde{\alpha}_0$ near the transition for the fully random ($w=\pi$) 2D symplectic model (Eq.~\ref{eq:sympham}) at energy $E=0$.
        The width of the transition $\Delta v$ is defined by where the curves cross the grey lines.
        Inset shows the scaling collapse.
    }\label{fig:randsympscaling}
\end{figure}

For the 2D model with tunable randomness, we show scaling data for the quasiperiodic case in Figure~\ref{fig:qpsympscaling} and the fully random case in Figure~\ref{fig:randsympscaling}.
Notice the strong finite size effects present in the quasiperiodic case (which suggests that corrections are needed to the simple scaling collapse form).
Rather than looking at the scaling collapse, we examine the width of the transition $\Delta v$, defined as the interval between where the curves cross $\tilde{\alpha}_0=2.2$ and $2.4$.  
This interval clearly contains the crossing point for the random case, but for the quasiperiodic case due to finite size effects the crossing point initially is outside of this interval and moving into it with increasing $L$.  
Nevertheless, the asymptotic scaling of the width $\Delta v\sim L^{-1/\nu}$ should be the same in the $L\rightarrow\infty$ limit as long as the width covers the scaling region of the curve (which it indeed does in both these cases).
This approach allows us to see how the scaling behavior changes with $L$, as discussed in the main text.


\begin{figure}[h]
    \centering
    \includegraphics[width=0.5\textwidth]{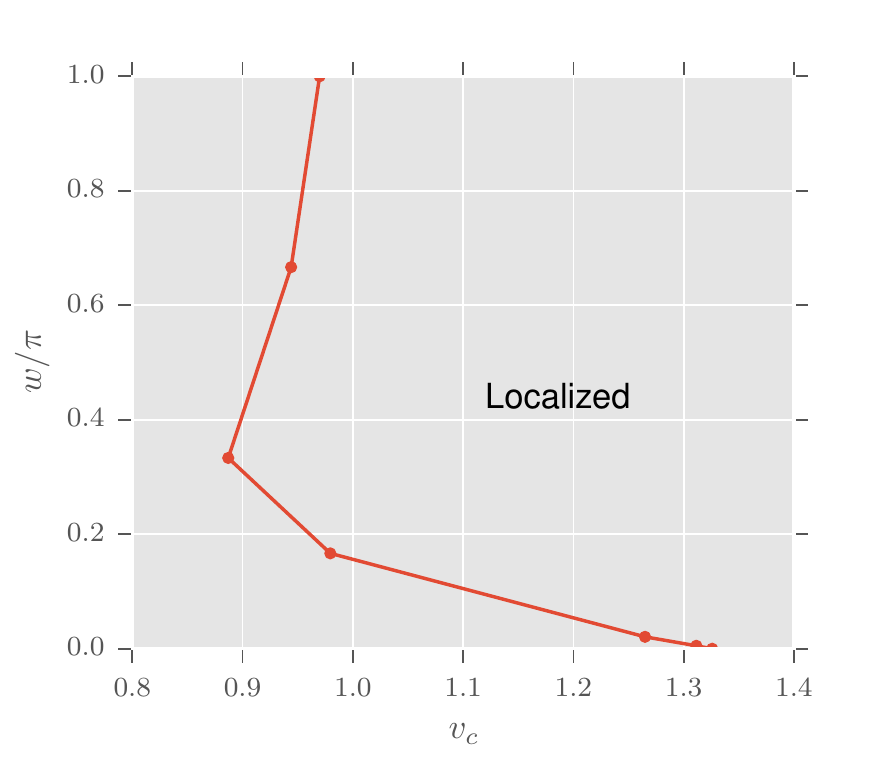}    
    \caption{The apparent phase diagram of the 2D symplectic Ando model, as a function of the potential strength $v$ and randomness $w$, coming from the finite-size scaling analysis.  This estimate of the localization transition shows a very strong dependence on randomness for weak randomness.}
    \label{fig:pd}
\end{figure}

As another probe of what is happening in this 2D symplectic Ando model as we turn on the randomness, we show the location of the apparent phase transition in Fig.~\ref{fig:pd}.  Here again we see results that are very sensitive to small added randomness.  Our preliminary interpretation of these results is that the randomness destabilizes an intermediate, possibly diffusive, phase in the nonrandom quasiperiodic model, so the localization phase boundary in the presence of randomness actually connects to the ballistic-to-intermediate phase transition in the quasiperiodic model.  The finite-size scaling analysis we did then is ``trying'' to instead connect this phase boundary to the intermediate-to-localized phase transition in the quasiperiodic model, thus producing a strong apparent crossover at small randomness.  In this scenario, the randomness localizes the intermediate phase and thus removes the intermediate-to-localized phase transition, while it converts the ballistic-to-intermediate phase transition in to the localization transition in the presence of the randomness. 

\end{document}